\documentclass[12pt]{article}

\usepackage{verbatim}
\usepackage{scicite}
\usepackage{times}
\usepackage[utf8]{inputenc}
\usepackage[T1]{fontenc}
\usepackage{booktabs}
\usepackage{wrapfig}
\usepackage{rotating}
\usepackage{pdflscape}
\usepackage{afterpage}
\usepackage{multirow}
\usepackage[utf8]{inputenc}
\usepackage{array}
\usepackage{makecell}
\usepackage{xurl}
\usepackage{xcolor}

\newenvironment{sciabstract}{%
\begin{quote} \bf}
{\end{quote}}

\title{The Potential and Perils of Generative Artificial Intelligence for Quality Improvement and Patient Safety}

\author
{Laleh Jalilian,$^{1}$ Daniel McDuff,$^{2}$ Achuta Kadambi$^{1}$\\
\\
\normalsize{$^{1}$University of California, Los Angeles (UCLA), Los Angeles, USA}\\
\normalsize{$^{2}$University of Washington, Seattle, USA}\\
\\
\normalsize{$^\ast$To whom correspondence should be addressed: ljalilian@mednet.ucla.edu.}
}

\date{}
\begin{document}

\baselineskip24pt

\maketitle

\begin{sciabstract}

Generative artificial intelligence (GenAI) has the potential to improve healthcare through automation that enhances the quality and safety of patient care. Powered by foundation models that have been pretrained and can generate complex content,
%and can be adapted to specific ``downstream'' tasks in a data efficient manner (i.e., with little or no finetuning), 
GenAI represents a paradigm shift away from the more traditional focus on task-specific classifiers that have dominated the AI landscape thus far. We posit that the imminent application of GenAI in healthcare will be through \emph{well-defined}, \emph{low risk}, \emph{high value}, and \emph{narrow} applications that automate healthcare workflows at the point of care using smaller foundation models. These models will be finetuned for different capabilities and application specific scenarios and will have the ability to provide medical explanations, reference evidence within a retrieval augmented framework and utilizing external tools. We contrast this with a general, all-purpose AI model for end-to-end clinical decision making that improves clinician performance, including safety-critical diagnostic tasks, which will require greater research prior to implementation. We consider areas where ``human in the loop'' Generative AI can improve healthcare quality and safety by automating mundane tasks. Using the principles of implementation science will be critical for integrating ``end to end'' GenAI systems that will be accepted by healthcare teams.

\end{sciabstract}

\section{Introduction}

Following the release of OpenAI's Chat Generative Pre-trained Transformer (ChatGPT) (San Francisco, CA, USA) in November 2022, the adoption of generative artificial intelligence (GenAI) and foundation models has witnessed a notable surge~\cite{OpenAI}. The impact was so dramatic that it contributed to attitudes changing towards how soon so-called Artificial General Intelligence (AGI) might be achieved~\cite{bengio2023managing}. At their most basic, foundation models receive text queries and generates text output in return~\cite{zhao2023survey, liang2023holistic}. In medical research settings, foundation models have been studied for use in various tasks such as operative and progress note-writing~\cite{gao2023progress,singh2023chatgpt}, answering patient questions~\cite{anastasio2023evaluating}, generating clinical summaries~\cite{singh2023chatgpt}, creating differential diagnoses~\cite{mcduff2023towards} and understanding and reasoning over personal health data~\cite{cosentino2024towards}. While the excitement surrounding the application of foundation models for improving clinician performance is palpable, these models have also demonstrated the ability to produce medically inaccurate or nonsensical outputs, hallucinate facts~\cite{ji2023survey,singhal2022large, lecler2023revolutionizing} and propogate existing biases ~\cite{omiye2023large}. Yet, despite these limitations, Generative AI is already being integrated into health systems~\cite{umeton2024gpt}. 

Generative AI holds the potential to transform many aspects of clinical medicine over the long term, ranging from improved patient diagnosis to precision treatment strategies. However, health systems are more likely to realize its immediate benefits by applying this technology towards improvements in quality, safety, and efficiency through well-defined, low-risk, high-value, and low-variation, repetitive and standardized use cases. A frequent approach in AI development for clinical applications is the pursuit of use cases which require near-perfect AI performance to achieve high-quality results~\cite{tu2024towards,saab2024capabilities}. While the focus on these use cases is indeed important, using foundation models in safety-critical areas like diagnostic and treatment recommendations will require considerable amount of research before safe implementation. Our experience with using foundation models has focused on understanding how it can be used for more straightforward tasks, where even average-quality outputs could elevate health system quality, safety, and efficiency at scale. In this Perspective piece, we consider areas where Generative AI may standardize care, minimize variability between providers, and provide value for over-burdened healthcare staff in clinical decision support tools at the point of care. 

%Ultimately, the challenge for all health systems will be in realizing the promise of these AI‐enabled innovations while overcoming their limitations. 

%This approach could accelerate the implementation of Generative AI in healthcare, while also accounting for potential constraints like training data availability and computational resources that health systems may face. 

%A recent report suggests that foundation models have the potential to tap into healthcare efficiency gains, potentially realizing savings of 5 to 10 percent from the estimated 1 trillion in potential improvements in U.S. healthcare spending.~\cite{NBERagra-2}, and generative AI may achieve these savings by prioritizing the development of use cases for administrative tasks.  

%Additionally, while the generalist approach is an important research direction for medicine, real world deployment may present stringent requirements for task-specific use cases. 

%We posit that initial penetration of GenAI into health systems could be through narrow, well-defined, low-risk, high-value applications using smaller foundation models that are optimized for specific use case scenarios.

\section{Foundation Models Capabilities}

A foundation model refers to a model that is (pre)trained on a large, broad dataset and can be adapted to a wide variety of downstream tasks without needing to be trained from scratch, giving it so-called "zero-shot" capabilities~\cite{bommasani2021opportunities}. Foundation models such as GPT~\cite{OpenAI}, BLOOM~\cite{workshop2022bloom}, LLaMA2~\cite{touvron2023llama} and StableDiffusion~\cite{rombach2021highresolution} are characterized by their scale in terms of training data, computational resources, and parameter size. Most foundation models utilize the transformer architectures~\cite{vaswani2017attention}, which can handle multi-modal data, including images, videos, audios, and 2D-waveforms. These models can be fine-tuned with smaller, more specific datasets to perform tasks that involve multiple modalities, such as image captioning, visual question-answering, or any task where understanding and generating content require both textual and non-textual information~\cite{li2023llava}. Fine-tuning updates the pretrained parameters to better suit the requirements of the specific task at hand. A key strength to the use of foundation models includes their ability to leverage their pretraining to adapt to tasks that were not explicitly part of their initial training data. They have demonstrated reasoning~\cite{wei2022chain} and planning~\cite{shen2024hugginggpt} capabilities and excel at data processing tasks such as integration and cleaning ~\cite{kayali2023chorus}.

\subsection{Retrieval-Augmented Generation} 

An important advancement in foundation models is the development of retrieval-augmented generation (RAG), which combines the generative capabilities of transformers with a retrieval mechanism that can access large external knowledge bases.  This allows the model to incorporate the latest data or retrieve and highlight the most pertinent information for specific use cases.  As an example, a RAG model could dynamically present the latest clinical guidelines~\cite{oniani2024enhancing} or patient-specific information in an interactive user interface. This ensures that the interface and generated text is not only based on the immediate data available but is also informed by relevant medical knowledge~\cite{gilbert2021generating}. Such a capability may augment clinical decision support systems that are dependent on current medical knowledge and limit the possibility of hallucinations. 
Training language models to utilize tools such as Internet search Application Programming Interfaces (APIs)~\cite{merrill2024transforming} is another method to retrieve information that the model might not be able to generate directly. 

\subsection{Larger Proprietary vs. Smaller Domain-Specific Models}

While larger proprietary and centralized models like GPT-3.5 first gained the attention of the public, these models can be costly, are associated with privacy risks, and may be inaccurate. Their use may also not be feasible for health systems that do not have HIPAA-compliant Azure instances.  Recent literature suggests that smaller models, when fine-tuned on specific domains, can surpass the performance of their proprietary counterparts~\cite{gupta2024prism}. This opens the possibility to use smaller, more efficient models that can address privacy, accuracy, and cost in healthcare.

\begin{figure*}[t]
    \centering
    \includegraphics[width=1\textwidth]{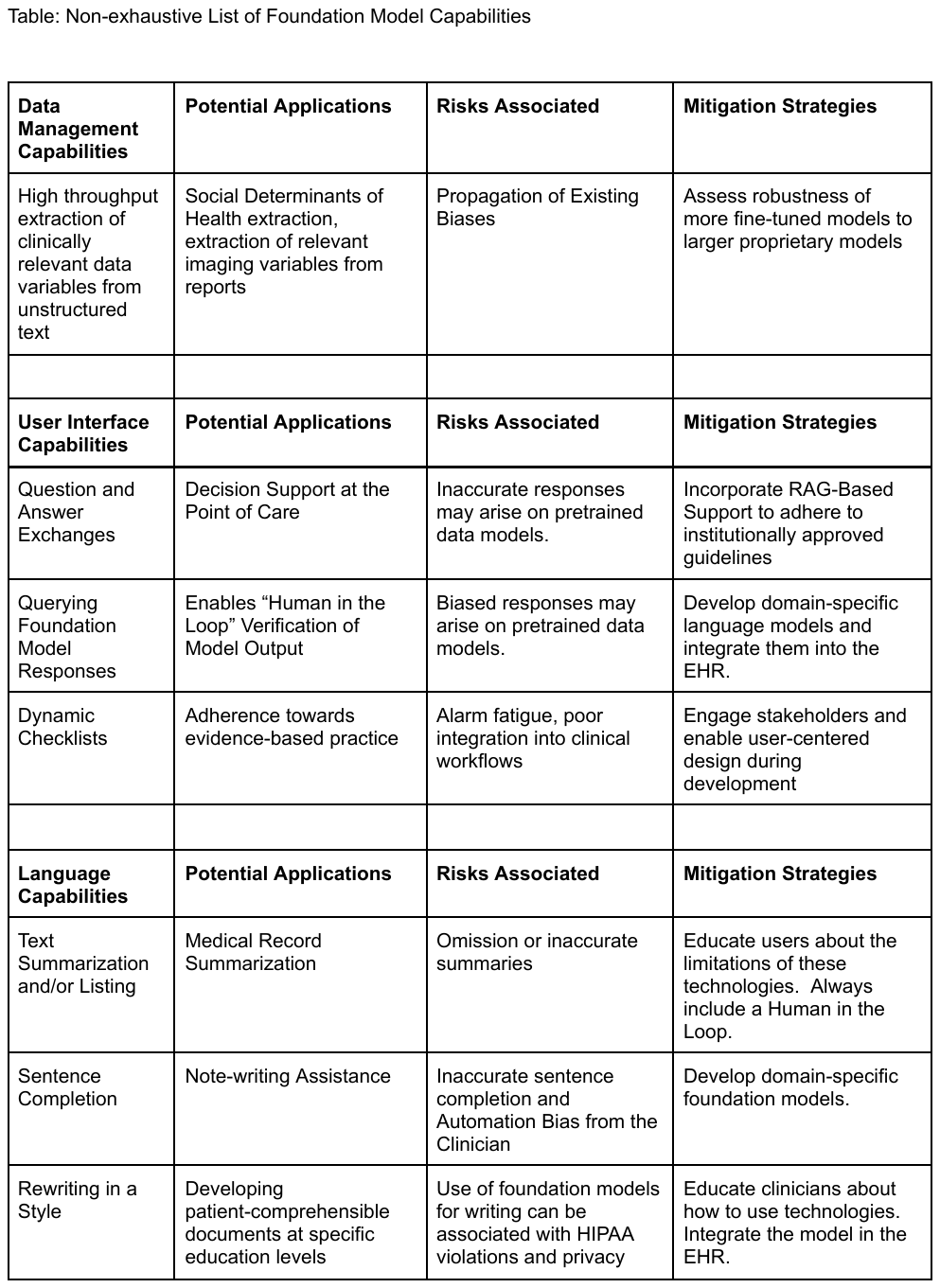}
    \caption{A Non-exhaustive List of Foundation Model Capabilities}
    \label{fig:example}
\end{figure*}

An optimal approach for foundation model-integrated application development will need to consider health system resource availability, domain specificity, degree of generalization, and the task-specific requirements. The usefulness and acceptance of the model will also depend on the quality of its output, and understanding when the model fails and the effects of failure~\cite{thomas2023revolutionizing}. Knowing these attributes will speed the development of GenAI use cases that support healthcare workers in engaging with the electronic health record (EHR) to complete tasks.

\section{Improving Quality and Safety at the Point of Care}

%Using GenAI for diagnostic tasks involves regulatory, clinical, and equity-related risks that require thorough research and risk mitigation before they can be safely implemented. 

Immediate applications of generative AI could prioritize non-diagnostic clinical and administrative tasks that can be labor-intensive, require manual review and input, and are well-defined~\cite{cowie2017electronic, casey2016using}. By automating lower risk tasks, generative AI could support healthcare workers and administrators in performing their duties more effectively.

%and aiding in the retrieval and synthesis of institutionally approved evidence-based care guidelines or other sources of information

%Current electronic prompts for evidence-based practice fail because they are agnostic to the specific situation, often firing when the practice in question is not indicated or is already being provided. As a result, they distract from the workflow without providing new information and leads to alarm fatigue, which in turn leads providers to ignore all prompts, even those that could help. Alternatively, generative AI applications can enhance the specificity of the alerts, triggering them when they are likely to be informative. As an example, a machine learning algorithm might predict that a patient will be unlikely to be assessed for delirium based on factors such as the patient’s characteristics, the bedside nurse's CAM-ICU assessment history, and the nursing:patient staffing ratio. Based on the predictive results, a resulting prompt for a CAM-ICU assessment, using Generative AI capabilities, would be rarer and thus be less likely to contribute to alarm fatigue, but also more actionable and increasing the impact of the alarm. 

\subsubsection{Dynamic User Interfaces to Support Interaction with Data for Informed Decision Making}

% support better workflow, clearer visual displays, RAG-supported display of institutionally accepted processes and evidence-based guidelines, and streamlined data entry to facilitate improved assessment, communication, and decision making. 

%Care teams that face challenges in accessing complex patient data may make uninformed or poor clinical decisions and subsequently may not write accurate notes to convey the patient's clinical status to other care teams, leading to clinical misinterpretation. 

%Additionally, while Generative AI can transform institutionally-approved evidence-based care guidelines into algorithms, which may then be linked to patient-level EHR data and used in DS through intelligent prompting. DS , but implementation of these protocols into actual clinical practice remains suboptimal~\cite{cheng2022treatment}. 

Current EHR interfaces present excessive data and are associated with adverse events, medical errors, and poor communication~\cite{jalilian2022next}. A nascent body of research focuses on using generative AI to create dynamic user interfaces~\cite{hu2019vizml} that facilitate healthcare workers interacting with and querying complex medical data, including patient data, compliance manuals, standard operating procedures, and evidence-based guidelines. Conceptually, this differs from training medicine-specific foundation models ~\cite{luo2022biogpt} or using models for prediction ~\cite{yang2022large}. Instead, generative AI-based systems may present an intuitive interface that can streamline interactions with multiple digital tools and sources of information, significantly augmenting clinicians’ capabilities by efficiently extracting and analyzing crucial information from vast amounts of patient data. Since the foundation model supplies knowledge from institutionally improved sources through RAG, the potential for hallucinations is limited. This enhances the practical impact of Generative AI tools while addressing current issues with using these models in clinical settings such as hallucinations~\cite{imrie2023redefining}. GenAI-supported dynamic interfaces can provide evidence-based information to healthcare teams at the point of care~\cite{kawamoto2005improving}, possibly improving adherence to evidence-based guidelines~\cite{cheng2022treatment}, reducing inter-provider variation, and provisioning the basic standard of care. Additionally, by transforming lengthy equipment manuals into interactive question and answer systems, GenAI can help clinical engineering teams with healthcare equipment maintenance while accessing key information as needed~\cite{jain2023generative}. This enhances engineer productivity and indirectly contributes to the equipment’s working quality.

\subsubsection{Improving Clinical Documentation for Clinicians and Administrators}

%Foundation models can significantly enhance notes by improving record accuracy and specifying precise medical temrniology

%For instance, the generation of clinical notes, which requires accuracy, context-awareness, and a deep understanding of medical terminology, can be significantly enhanced by these models.

The accurate and comprehensive documentation of clinical notes is paramount for ensuring effective communication, care quality, and patient safety. As they currently exist, notes suffer from issues such as note bloat (copying and pasting), information overload (automatic inclusion of data and administrative documentation), and disorganization. Generative AI's assistive writing abilities can reduce documentation burden for clinicians by using language capabilities to document specific medical terminology and increase record accuracy. This would this provide higher quality documentation and training data for future iterations of EHR foundation models~\cite{castaldi2019introducing}. They can assist clinical documentation by writing referral notes and prescriptions and generate concise summaries~\cite{psotka2020streamlining, shah2023creation}, among other applications. They can parse medical records and synthesize them into clear, concise summaries of active and historical conditions. They can also interact with medical records by converting unstructured data into structured formats. Recent work demonstrates that foundation models can extract key data elements from patient notes and reformat this information into a structured layout to aid in filtering clinical trials~\cite{gupta2024prism}. Additionally, a "prompt engineering" framework has enabled batch queries to process large volumes of pathology reports for structured information extraction and estimation without requiring extensive task-specific human annotation and model training~\cite{huang4488945critical}. In complex, high-stakes online pharmacy prescription processing, where the magnitude of prescriptions received daily can contain many inaccuracies, a data-driven pharmacy support system using Generative AI has already enhanced quality and reduced errors, allowing pharmacy teams to focus on ensuring patient safety and well-being ~\cite{pais2024large}.

\subsubsection{Dynamic Checklists for Adherence to Evidence-Based Practice}

Since gaining traction in the context of perioperative care, use of checklists in routine patient care has expanded to multiple fields of medicine. Unlike static checklists, dynamic checklists display data appropriate to a patient’s clinical context but also provide guidelines relevant to the checklist items. Dynamic checklists significantly improve compliance with best practices during ward rounds in the ICUs~\cite{Bie2017Intelligent} and are associated with reductions in length of stay~\cite{Bie2020Intelligent}. They are supported by hard-coded logic, but generative AI can be applied to assess the patient’s eligibility for specific practices, emphasize items that are most important, and remove items that aren't indicated, to ensure that teams are compliant with best practices, all in a user-friendly interface. They can be used during rounding, during note writing, or during procedural care to ensure that all aspects of care are addressed. 

%A key area of study should be on assessing clinician perception on the efficiency and effectiveness of care from the use of dynamic checklists, with assessments on key outcome measures including cognitive load, work load, and time load.

A concrete example of where generative AI can be applied to dynamic checklists is adherence to lung protective ventilation (LPV) in patients on mechanical ventilators, whether intra-operatively or in the ICU. In ICUs that adhere to LPV, noticeable improvements in both short- and long-term patient outcomes and reductions in healthcare costs~\cite{pun2019caring, barnes2017improving} are noted. However, LPV is not universally implemented used across ICUs and ORs, because of challenges including developing and digitizing clinical protocols for LPV management and a lack of IT resources to acquire and analyze LPV metrics within the EHR~\cite{costa2017identifying}. In the OR and ICUs of the future, a generative AI-enhanced dynamic checklist integrated within the EHR can assess ideal body weight, tidal volumes on the ventilator, and adherence to LPV protocols.  This can help to standardize operative and ICU ventilator management, reduce inter-provider practice variation, and ensure that any patient on a ventilator has received consideration towards applying LPV. 

\subsubsection{Health System Data Auditing, Extraction, and Population}

Foundation models could become a core component of data discovery, exploration, and extraction systems in healthcare. They have demonstrated superior capability on three data discovery tasks, including table-class detection, column-type annotation and join-column prediction~\cite{kayali2023chorus}. They could be applied to extracting data from disparate data warehouses and data lakes in health systems to reduce data silo'ing, which limits the ability of health systems to use their data effectively. Currently, notable reasons for silo'ing include different departments using different data formats, which impacts data aggregation in one central area.  Examples of this include perioperative data warehouses which hold data captured in EHR systems but may not include data from other relevant systems, such as imaging systems. Foundation models can help in extracting information from disparate data sources and formats and can be used to translate different medical terminologies and data structures into a unified format, facilitating easier data integration into individual department-level data warehouses.  This could increase the utility of health system data and improve the ability of health systems to engage in predictive analytics or even basic auditing. 

\section{Implementation Path for Responsible Generative AI}

While a significant body of scholarship has focused on model development, realizing generative AI’s potential in healthcare requires structured and context-sensitive approaches rooted in implementation science~\cite{bauer2020implementation}. Multidiscplinary teams consisting of experts in computer science, healthcare workers, administrative staff, and other members of the healthcare community will need to guide the development of "end to end" systems, with consideration into the details around design, user experience, workflow integration, ease of adoption, and continued performance monitoring within clinical workflows. Ideally, these models should function within interactive systems that are usable, support workflow, and meet the needs of end users~\cite{searl2010time, lyon2016user, schwartz2021clinician}. Studies will be needed to quantitatively assess how GenAI systems impact user experience, efficiency, productivity, and workload~\cite{friedberg2014factors, nazar2021systematic, rajawat2021robotic,esmaeilzadeh2021patients} in real world clinical workflows, in order to continuously verify the purported benefits of the technology. The total cost of implementation, which includes not only the cost to run the model but also expenses associated with monitoring, maintenance, and any necessary infrastructure adjustments, will also need to be estimated.

%\begin{figure*}[t]
%    \centering
%    \includegraphics[width=1\textwidth]{figures/Output Quality vs Task Complexity (1).pdf}
%    \caption{GenAI Use Case Development should focus on understanding the capabilities of the foundation model, the specific task complexity it is asked to execute, and the quality of its outputs. Classifying foundation model capabilities within a complexity-output quality matrix could serve as guidance for health systems who wish to use foundation models to improve the support of specific tasks and may help to different users’ needs and expectations.}
%    \label{fig:example}
%\end{figure*}

%\begin{figure*}[t]
%    \centering
%    \includegraphics[width=1\textwidth]{images/***}
%    \caption{Panel A: Framework for GenAI Use Case Development that focuses on understanding the capabilities of the foundation model, the specific task complexity it is asked to execute, and the quality of its outputs. Panel B: Classifying foundation model capabilities within a complexity-output quality matrix could serve as guidance for health systems improving the support of specific tasks. Furthermore, for developing assistive tools that can adapt to users’ expertise and context, a complexity-output quality matrix might help determine which capabilities best support different users’ needs and expectations.}
%    \label{fig:example}
%\end{figure*}

\subsection{Risks}

Risks with foundation models for any task include the potential for hallucinations, omission, and bias~\cite{webb2022addressing}. A major challenge in deploying foundation models for healthcare tasks is how to control the output to prevent hallucinations. Current models do not evaluate the quality or provide a measure of uncertainty for their outputs. Additionally, many datasets contain too much data to provide to a foundation model in a prompt, which may lead to outputs with clinically relevant omissions. The quality of data used in both pre-training and instruction fine tuning of foundation models for different tasks impacts the performance of the model, and this data has to be carefully assessed for quality, quantity, and diversity~\cite{de2022guidelines}. %The accuracy and completeness of EHR data vary are influenced by factors such as data entry practices~\cite{Zakim2021Computerized}.  

Algorithmic bias towards underrepresented minorities may perpetuate or even increase healthcare disparities~\cite{lucy2021gender, abid2021large, kadambi2021achieving, vilesov2022blending}, and techniques will be needed to mitigate these biases to ensure that LLMs promote health equity across diverse populations~\cite{wang2022synthetic, mcduff2022scamps}. If the training data contains under-representation of certain demographics groups or outdated medical practices, generative AI may inadvertently embed theses biases in its outputs and potentially cause harm. These risks threaten acceptance and adoption of any future AI tools for use in healthcare ~\cite{zou2021ensuring}. It is also not yet well understood if GenAI may exert unpredictable effects on the quality and patient safety, and measurement of various outcomes such as fairness and safety in the specific environment will need to be done to assess the impact of the technology. Finally, particular attention must be given to the study of automation complacency~\cite{parasuraman2010complacency}.

\section{Conclusion}

%Healthcare systems worldwide face crises of inconsistent quality, high inter-provider variation that endangers patient safety~\cite{reddy2020artificial}, and clinician burnout.  

Early penetration of Generative AI into healthcare may be on tasks where it performs reasonably well in ways that enable digital standardization towards delivering higher quality, more efficient, and safer care.  As healthcare slowly integrates GenAI, using principles of implementation science and studying the impact on patient outcomes and clinical operations will be important. A careful and coordinated approach that involves scrutiny, customization of this technology to its specific context, and regularly verifying how GenAI assists in clinical encounters, could address concerns associated with adopting Gen AI in healthcare. While considerable amount of work still needs to be done to ensure that "human in the loop" GenAI systems are practical, reliable, and safe, we envision a future where GenAI systems can meaningfully accelerate quality, safety, and efficiency in healthcare delivery.

%We offer a conceptual framework that may allow for Generative AI to be implemented in healthcare settings, where its capabilities, task complexity, and output quality are considered when designing "human in the loop" generative AI systems.

%Third, regularly verifying (and reverifying) the presumed benefits of using such technology assistance via structured assessments and in real-time clinical encounters.

\bibliography{bibliography}
\bibliographystyle{Science}
\end{document}